\documentclass{elsart}
\usepackage{psfig,latexsym,amssymb}

\newcommand{\ba}{\begin{array}} \newcommand{\ea}{\end{array}}
\renewcommand{\thesection}{\arabic{section}.}
\renewcommand{\theequation}{\arabic{section}.\arabic{equation}}

\begin{document}

\begin{frontmatter}

\title{Exact solution for the Bariev model
with boundary fields}
\author[PortoAlegre]{A.~Foerster},
\author[PortoAlegre]{X.-W.~Guan},
\author[Queensland]{J. ~Links},
\author[RiodeJaneiro]{I.~Roditi},
\author[Queensland]{H.-Q~Zhou},
\address[PortoAlegre]{Instituto de Fisica da UFRGS,
                     Av.\ Bento Goncalves, 9500,\\
                     Porto Alegre, 91501-970, Brasil}
\address[RiodeJaneiro]{Centro Brasileiro de Pesquisas Fisicas,\\Rua Dr. 
Xavier Sigaud 150, 22290-180, Rio de Janeiro- RJ, Brasil}
\address[Queensland]{Centre for Mathematical Physics, 
Department of Mathematics, \\ The University of Queensland, 4072, Australia}

\begin{abstract}
The Bariev model with open boundary conditions is introduced and
analysed in detail in the framework of the Quantum Inverse Scattering
Method. Two classes of independent boundary reflecting $K$-matrices
leading to four different types of boundary fields are obtained by solving
the reflection equations.
The models are exactly solved by means of the algebraic nested
Bethe ansatz method and the four sets of Bethe ansatz equations as
well as their corresponding energy expressions are derived.

\end{abstract}

\end{frontmatter}

\section{Introduction}
\label{sec1}
Much work has been devoted, during the last years, towards a better 
understanding of integrable models of strongly correlated electrons. 
Such an interest arises from the efforts to unveil a theory that can
provide a consistent picture for observed phenomena of condensed matter
physics such as high-$T_c$ 
superconductivity. Two prototypes of integrable models for this purpose
are the one-dimensional (1D) Hubbard model solved by Lieb and 
Wu \cite{LW} (see also \cite{reprints}) and the t-J model \cite{azr}
(which is integrable for the {\it supersymmetric} coupling \cite{ffk}), 
mainly because of their relevance in the description of  electronic mechanisms. 
This is  justified on physical grounds since the electron hopping is 
strongly disturbed by the on-site Coulomb interaction in the 1D Hubbard model and by the 
spin fluctuations through the antiferromagnetic coupling for the supersymmetric
t-J model. More recently, motivated by the inclusion of additional
interactions whether through internal impurities or external boundary fields, 
a renewed interest in the study of integrable 
strongly correlated electron systems has 
taken place. Some examples of the latter are the 1D Hubbard open chain 
\cite{OPC4}, the supersymmetric t-J model with boundary fields \cite{t-J1,t-J2} 
and other integrable electronic systems with boundary interactions 
associated with 
Lie superalgebra symmetries \cite{Lie-sup}. 

The study of integrable systems with open boundary conditions (OBC) 
\cite{OPC4,OPC1,OPC2,OPC3} has turned into an active domain since the 
introduction by Sklyanin \cite{EK} of a generalization of the Quantum Inverse 
Scattering Method (QISM) \cite{QISM1,QISM2} which provides a systematic 
approach to handle the boundary problem by introducing a set of equations, 
called reflection equations (RE). The  solutions of these equations,
referred to as  the boundary 
K-matrices, in turn introduce boundary interactions into the Hamiltonian
of the system in such  a manner that integrability is preserved. 
Indeed, the presence of boundary fields drastically changes surface
critical properties, and in some occasions results in the appearance of
so-called boundary bound states \cite{b-eff1,b-eff2,b-eff3}. Nevertheless, 
in spite of these results, several models still lack a comprehensive 
study regarding OBC, one of those being the 1D Bariev model \cite{Bar1,Bar2}. 

The 1D Bariev (interacting $XY$) chain is also a Hubbard-like integrable 
model of special interest, as it exhibits the existence of hole pairs of 
Cooper type which are relevant to theories of superconductors. 
The integrability 
of the 1D Bariev model within the framework of the QISM was first
established  by Zhou in \cite{zhou} (see also \cite {wadd}). To achieve this 
goal 
one has to build non-trivial higher conserved currents that allow the 
identification of a quantum $R$-matrix satisfying the Yang-Baxter 
equation \cite{QISM1,QISM2}. This $R$-matrix  
does not possess the difference property for the spectral 
parameter, a feature that is also shared by the $R$-matrix associated
with the Hubbard model \cite{shastry}. 
Soon after this parametrization of the model had 
been obtained, the algebraic Bethe ansatz solution for the periodic
chain was calculated independently by Martins and Ramos \cite{Mar} and 
by Zhou \cite{zhou2}. On the other hand, as far as we know, 
studies concerning OBC for the Bariev model appear incomplete. 
This can be understood in light of the fact that the $R$-matrix is
presented in a complicated form with the absence of both additivity for the
spectral parameter and crossing unitarity. As such, we must resort to
searching for solutions of a  
much more intricate RE. Though some known solutions to the RE for the 
1D Bariev 
open chain provide a certain class of boundary K-matrices \cite{zhou3}, 
a thorough investigation is still lacking and, moreover, 
the algebraic Bethe ansatz solution has not been obtained yet.

In this paper, we shall study in detail the OBC for the 1D Bariev 
chain within the framework of the QISM. 
We will show that
besides the $K$-matrices obtained in \cite{zhou3} there exists a new class
of solutions of the RE which leads to pure magnetic boundary
fields in the Hamiltonian and provides more general nesting boundary
$K$-matrices for a hidden asymmetric $XXZ$ open chain. 
Here it is worth noticing that this class of boundary fields may 
have a feasible realization by applying
boundary external voltages in experiments on quantum wires \cite{QW}. 
By means of the algebraic Bethe ansatz technique, we obtain the eigenvalues
of the
transfer matrix of the 1D Bariev model for  four different kinds of possible
boundary fields. For all cases we present the Bethe ansatz 
equations and the energy spectrum in explicit form, which provides
a first step towards investigating
the surface critical properties for the 1D Bariev
model with these different boundary fields.

The paper is organized as follows. In section 2 we construct the
Bariev model with four kinds of boundary fields by means of the
QISM adapted to special boundary conditions. The basic quantities,
e.g., the $R$-matrix, the matrices $K_{\pm}$ defining the
boundaries, the monodromy matrices and the
transfer matrices are defined.
In particular, we solve the RE and obtain two independent
classes of solutions for the boundary
$K$-matrices using the variable-separation method \cite{MG}.
The relation between the transfer matrices and the Bariev
Hamiltonians with different boundary fields is established.
Sections 3 and 4 are devoted to the solution of the models 
through Bethe ansatz methods. In particular, in section 3
we discuss the action of the transfer matrix on the
pseudovacuum state. In section 4 we perform the nested algebraic
Bethe ansatz and find the Bethe ansatz equations, the eigenvalues of the
transfer matrices and the energy spectrum of the system.  Section 5
is reserved for our conclusions and discussions.

\section{The Bariev open chain with boundary fields}
\label{sec2}
Let us begin by introducing the Bariev model with
boundary fields whose Hamiltonian reads
\begin{eqnarray}
H & = & \sum_{j=1}^{N-1}\left\{\left(\sigma ^{+}_{j}\sigma ^{-}_{j+1}
+\sigma^{-}_{j}\sigma^{+}_{j+1}\right)\exp
\left(\beta\tau ^{+}_{j+1}\tau^{-}_{j+1}\right)\right.\nonumber\\
& &
\left.+\left(\tau ^{+}_{j}\tau^{-}_{j+1}+\tau_{j}^{-}\tau^{+}_{j+1}\right)
\exp\left(\beta\sigma^{+}_{j}\sigma^{-}_{j}\right)\right\}
+{\rm b.t.} \label{Ham+bt}
\end{eqnarray}
where the four different boundary terms are given by
\begin{eqnarray}
{\rm b.t.}^{({\rm I},{\rm I})} & = &
-\frac{\exp\left(\beta \right)}{c_2}\left[\sinh\beta\,\sigma^{z}_{N}
\tau^{z}_{N}+\cosh\beta \,\left(\sigma ^{z}_{N}+\tau ^{z}_{N}\right)\right]
\nonumber\\
& &
-\frac{\exp\left(\beta \right)}{c_1}\left[\sinh\beta\,\sigma^{z}_{1}
\tau^{z}_{1}+\cosh\beta \,\left(\sigma ^{z}_{1}+\tau ^{z}_{1}\right)\right],
\label{11}\\
{\rm b.t.}^{({\rm II},{\rm II})} & = &
\frac{1}{2}\exp\left(\beta \right)c_2\sigma ^{z}_{N}+\frac{1}{2}
\exp\left(\beta \right)c_1\tau ^{z}_{1},
\label{22}
\\
{\rm b.t.}^{({\rm I},{\rm II})} & = & -\frac{\exp\left(\beta \right)}{c_2}
\left[\sinh\beta\,\sigma^{z}_{N}\tau^{z}_{N}+\cosh\beta \,\left(\sigma ^{z}_{N}
+\tau ^{z}_{N}\right)\right]\nonumber\\
& &+\frac{1}{2}\exp\left(\beta \right)c_1\tau ^{z}_{1},\label{12}\\
{\rm b.t.}^{({\rm II},{\rm I})} & = & \frac{1}{2}\exp\left(\beta 
\right)c_2\sigma ^{z}_{N}\nonumber\\
& &-\frac{\exp\left(\beta \right)}{c_1}\left[\sinh\beta\,\sigma^{z}_{1}\tau^{z}_{1}+\cosh\beta 
\,\left(\sigma ^{z}_{1}+\tau ^{z}_{1}\right)\right].
\label{21}\end{eqnarray}
Above $\sigma _{i}$ and $\tau _{i}$ are the two commuting species of
Pauli matrices acting on site $i$, $\beta $ is a coupling constant and
$c_{1},\,c_{2}$ are the boundary parameters characterizing the strength of
the boundary fields. Notice that in cases (\ref{11},\ref{22})
the boundary terms at the left and right ends are consistent with the
bulk symmetry which is a combination of the inversion $j \rightarrow
N-j+1$ and the exchange $\sigma \leftrightarrow \tau$. The cases
(\ref{12},\ref{21}) are interchanged under this symmetry.

In order to derive these models, we first consider the following
$R$-matrix \cite{zhou}
$$R(u,v)=$$
\begin{equation}
\left(
\matrix{
\rho _1&0&0&0&0&0&0&0&0&0&0&0&0&0&0&0\cr
0&\rho _{2}&0&0&\rho _3&0&0&0&0&0&0&0&0&0&0&0\cr
0&0&\rho _{2}&0&0&0&0&0&\rho _3&0&0&0&0&0&0&0\cr
0&0&0&\rho _4&0&0&\rho _5&0&0&\rho _6&0&0&\rho _9&0&0&0\cr
0&\rho _{3}&0&0&\rho _2&0&0&0&0&0&0&0&0&0&0&0\cr
0&0&0&0&0&\rho _1&0&0&0&0&0&0&0&0&0&0\cr
0&0&0&\rho _{12}&0&0&\rho _7&0&0&\rho _{15}&0&0&\rho _5&0&0&0\cr
0&0&0&0&0&0&0&\rho _9&0&0&0&0&0&\rho _{11}&0&0\cr
0&0&\rho _{3}&0&0&0&0&0&\rho _2&0&0&0&0&0&0&0\cr
0&0&0&\rho _{13}&0&0&\rho _{15}&0&0&\rho _{10}&0&0&\rho _6&0&0&0\cr
0&0&0&0&0&0&0&0&0&0&\rho _1&0&0&0&0&0\cr
0&0&0&0&0&0&0&0&0&0&0&\rho _8&0&0&\rho _{11}&0\cr
0&0&0&\rho _{14}&0&0&\rho _{12}&0&0&\rho _{13}&0&0&\rho _4&0&0&0\cr
0&0&0&0&0&0&0&\rho _{11}&0&0&0&0&0&\rho _{8}&0&0\cr
0&0&0&0&0&0&0&0&0&0&0&\rho _{11}&0&0&\rho _{8}&0\cr
0&0&0&0&0&0&0&0&0&0&0&0&0&0&0&\rho _1\cr }\right)\label{R}
\end{equation}
whose elements (Boltzmann weights) are given
explicitly in Appendix A.

It satisfies the non-additive Yang-Baxter equation
\begin{equation}
R_{12}(u,v) R_{13}(u,w) R_{23}(v,w) =
R_{23}(v,w) R_{13}(u,w) R_{12}(u,v)
\end{equation}
and enjoys the following properties
\begin{eqnarray}
R_{12}(u,v)R_{21}(v,u) & = & 1,\\
\tilde{R}_{21}^{{\rm t}_1}(-v,u)R_{12}^{{\rm t}_2}(u,-v) & = & 1,\label{tildeR1}\\
\tilde{R}_{12}^{{\rm t}_2}(-u,v)R_{21}^{{\rm t}_1}(v,-u) & = & 1.\label{tildeR2}
\end{eqnarray}
Above, ${\rm t}_i $ denotes matrix transposition in the $i$th space
and $R_{21} = P.R_{12}.P$, where $P$ is the permutation operator. 
The matrix $\tilde{R}$ is a new object introduced in Ref. \cite{zhou3}
in order to handle integrable boundary terms. We may take either
(\ref{tildeR1}) or (\ref{tildeR2}) as its definition.
We remark here that neither the crossing unitarity nor
the difference property is valid for this $R$-matrix, which
turns the calculations into a complex problem.

The Yang-Baxter algebra can also be verified for the monodromy matrix $T$
\begin{equation}
R_{12}(u,v)\stackrel{1}{T}(u)\stackrel{2}{T}(v)=
\stackrel{2}{T}(v)\stackrel{1}{T}(u)R_{12}(u,v),
\label{2c}
\end{equation}
which is defined, as usual, by
\begin{equation}
T(u)=L_{0N}(u)\cdots L_{01}(u).  \label{TM}
\end{equation}
Above the subscripts $0$ and $N, \dots, 2,1$ denote the
auxiliary and quantum spaces, respectively
and the local Lax operator is given by
\begin{equation}
L_{j}(u)=\left(\matrix{P^{+}_{j}Q^{+}_{j}&f_1(u)P^{+}_{j}\tau^{-}_{j}&f_1(u)\sigma^{-}_{j}\tilde
{Q}^{+}_{j}&f_1(u)f_2(u)\sigma^{-}_{j}\tau^{-}_{j}\cr
f_1(u)\tilde{P}^{+}_{j}\tau^{+}_{j}&\tilde{P}^{+}_{j}Q^{-}_{j}&f^2_2(u)\sigma^{-}_{j}\tau^{+}_{j
}&f_2(u)\sigma^{-}_{j}\tilde{Q}^{-}_{j}\cr
f_1(u)\sigma^{+}_{j}Q^{+}_{j}&f_1^2(u)\sigma 
^{+}_{j}\tau_{j}^{-}&P^{-}_{j}\tilde{Q}^{+}_{j}&f_2(u)P^{-}_{j}\tau_{j}^{-}\cr
f_1(u)f_2(u)\sigma^{+}_{j}\tau^{+}_{j}&f_2(u)\sigma^{+}_{j}Q^{-}_{j}&f_2(u)\tilde{P}^{-}_{j}\tau
_{j}^{+}&\tilde{P}_{j}\tilde{Q}^{-}\cr }\right)  \label{Lax}
\end{equation}
where
\begin{eqnarray}
P_{j}^{\pm} & = & \frac{1}{2}(1\pm \sigma ^{z}_{j})+\frac{u}{2}\exp\left(\beta \right)(1\mp 
\sigma^{z}_{j});\nonumber\\
\tilde{P}_{j}^{\pm} & = & \frac{1}{2}(1\pm \sigma ^{z}_{j})+\frac{u}{2}(1\mp 
\sigma^{z}_{j});\nonumber\\
Q_{j}^{\pm} & = & \frac{1}{2}(1\pm \tau ^{z}_{j})+\frac{u}{2}\exp\left(\beta \right)(1\mp 
\tau^{z}_{j});\nonumber\\
\tilde{Q}_{j}^{\pm} & = & \frac{1}{2}(1\pm \tau ^{z}_{j})+\frac{u}{2}(1\mp 
\tau^{z}_{j});\nonumber\\
f_1(u) & = &\sqrt{1+u^2\exp\left(2\beta \right)},\,\,f_2(u)=\sqrt{1+u^2}\nonumber
\end{eqnarray}
Next, following the QISM adapted to the case of special boundary conditions,
we define the doubled monodromy matrix as
\begin{equation}
U_{-}(u)= T(u)K_{-}(u)T^{-1}(-u),
\label{DMM}
\end{equation}
such that the transfer matrix is given by
\begin{equation}
\tau (u)=Tr_{0}K_{+} (u)U_{-}(u)
\label{DTM}
\end{equation}
Above $T^{-1}$ is the inverse of the monodromy $T$ and
$K_{\pm}$ are the matrices defining the boundaries. The requirement
that they obey the RE
\cite{zhou3}
\begin{eqnarray}
& &R_{12}(u,v)\stackrel{1}{K_-}(u)R_{21}(v,-u)\stackrel{2}{K_-}(v)\nonumber\\
& &= \stackrel{2}{K_-}(v)R_{12}(u,-v)\stackrel{1}{K_-}(u)R_{21}(-v,-u),  \label{RE1}\\
& &R_{21}^{{\rm t}_1{\rm t}_2}(v,u)\stackrel{1}{K_+^{
{\rm t}_1}}(u)\tilde{R}_{12}(-u,v)
\stackrel{2}{K_+^{{\rm t}_2}}(v)=  \nonumber \\
& &\stackrel{2}{K_+^{{\rm t}_2}}(v)\tilde{R}_{21}(-v,u)
\stackrel{1}{K_+^{{\rm t}_1}}(u) R
_{12}^{{\rm t}_1{\rm t}_2}(-u,-v)  \label{RE2}
\end{eqnarray}
together with the Yang-Baxter algebra assure that the transfer matrix
commutes for different spectral parameters, proving the integrability
of the model.
Therefore the transfer matrix (\ref{DTM}) may be considered as the
generating fuction of infinitely many  integrals of motion for the system.
We emphasize that there is no isomorphism between
the matrices $K_{+}(u)$ and $K_{-}(u)$,
due to the absence of the crossing unitarity
for the $R$-matrix.
Therefore, we have to solve
the two RE separately in order to fix the boundaries.
In Appendix A, this calculation is presented in detail
using the {\it variable-separation} prescription.
This approach, which was already proposed in \cite{MG}
to treat the $D^{(2)}_2$ model with OBC, is extended
here for the case of an $R$-matrix without the
difference property.
It is in fact a systematic method which simplifies
drastically the calculation of the
$K_{\pm}$ matrices.
After a tedious algebra, we find two different classes of 
boundary $K_\pm $ matrices
consistent with the integrability of the Hamiltonian (\ref{Ham+bt}),
listed below
\begin{equation}
K_{\pm }(u)=\left(\matrix{K1_{\pm }(u)&0&0&0\cr 0&K2_{\pm }(u)&0&0\cr
0&0&K3_{\pm }(u)&0\cr 0&0&0&K4_{\pm }(u)\cr}\right),  \label{BK}
\end{equation}
Solutions for $K_{-}$, corresponding to the left boundary \\
Case I
\begin{eqnarray}
K1_{-}(u) & = &(u-c_1)(h^2u-c_1),\nonumber\\
K2_{-}(u) & = &(c_1-u)(h^2u+c_1),\nonumber\\
K3_{-}(u) & = &(c_1-u)(h^2u+c_1),\label{Km1}\\
K4_{-}(u) & = &(u+c_1)(h^2u+c_1),\nonumber
\end{eqnarray}
Case II
\begin{eqnarray}
K1_{-}(u) & = &(1+c_1hu),\nonumber\\
K2_{-}(u) & = &(1-c_1hu),\nonumber\\
K3_{-}(u) & = &(1+c_1hu),\label{Km2}\\
K4_{-}(u) & = &(1-c_1hu).\nonumber
\end{eqnarray}
Solutions for $K_{+}$, corresponding to the right boundary \\
Case I
\begin{eqnarray}
K1_{+}(u) & = &(c_2u+1)(h^2c_2u+1),\nonumber\\
K2_{+}(u) & = &h^2(c_2u-1)(c_2u+1),\nonumber\\
K3_{+}(u) & = &(c_2u-1)(c_2u+1),\label{Kp1}\\
K4_{+}(u) & = &(c_2u-1)(c_2u-h^2),\nonumber
\end{eqnarray}
Case II
\begin{eqnarray}
K1_{+}(u) & = &h(hu-c_2),\nonumber\\
K2_{+}(u) & = &h(hu-c_2),\nonumber\\
K3_{+}(u) & = &(u+hc_2),\label{Kp2}\\
K4_{+}(u) & = &(u+hc_2)\nonumber
\end{eqnarray}
where $h=\exp\left(\beta\right)$. These two classes of boundary $K_{\pm
}$-matrices provide four possible choices of BC, according to the
combination of the boundary
pairs ($K_{+}(u),K_{-}(u)$), which are being labeled by (I,I), (II, II),
(I,II) and (II,I) and originate the boundary terms ${\rm
b.t.}^{({\rm I},{\rm I})}$, ${\rm b.t.}^{({\rm II},{\rm II})}$, ${\rm
b.t.}^{({\rm I},{\rm II})}$ and ${\rm b.t.}^{({\rm II},{\rm I})}$
respectively.  Taking into account the fact that $TrK_{+}(0)=0$ and
$TrK_{+}^{'}(0)=0$ in the cases (I,I) and (I,II) and $TrK_{+}(0)=0$ in the
cases (II,II) and (II,I), we can show that the Hamiltonian (\ref{Ham+bt})
with the boundary terms ${\rm b.t.}^{({\rm I},{\rm I})}$ and ${\rm
b.t.}^{({\rm I},{\rm II})}$ is related to the transfer matrix
matrix (\ref{DTM}) in the following way
\begin{equation}
\tau (u)=C_{1}u^2+C_{2}(H+{\rm const.})~u^{3}+\cdots,  \label{H-DTM-1}
\end{equation}
while
\begin{equation}
\tau (u)=C_{3}u+C_{4}(H+{\rm const.})~u^{2}+\cdots   \label{H-DTM-2}
\end{equation}
for the Hamiltonian (\ref{Ham+bt}) with the boundary
terms ${\rm b.t.}^{({\rm II},{\rm II})}$ and ${\rm b.t.}^{({\rm II},{\rm I})}$.
The difference in the leading term expansions
(\ref{H-DTM-1},\ref{H-DTM-2}) is a consequence of the fact that for
case I the matrix $K_+(u)$ is quadratic in the variable $u$ while for
case II it is linear.
Above $C_{i},\,i=1,\cdots 4,$ are some scalar functions of the boundary
parameters.
In the next sections, we shall be focusing in the solution of the
eigenvalue problem of
the transfer matrix (\ref{DTM}) and, consequently, in the diagonalization
of the present model.
\section{Algebraic Bethe ansatz I: Action of the operators on the pseudovacuum}
\label{sec3}

In order to carry out the algebraic Bethe ansatz for the
Bariev model with boundaries, we first need to find the
eigenvalues and eigenvectors of the transfer matrix (\ref{DTM})
\begin{equation}
\tau |\Phi_n\rangle= \lambda |\Phi_n\rangle
\label{eigenv}
\end{equation}
This will be achieved by extending the Bethe ansatz
techniques developed in \cite{MR},\cite{GJPA} for
the Hubbard model to the
present case.
Following this prescription, the eigenvectors of the
transfer matrix are given by
\begin{equation}
|\Phi_n\rangle=
{\bf \Phi_{n}}.{\bf F}|0\rangle .
\label{np-state}
\end{equation}
where the components of $ {\bf F}$ are coefficients
of an arbitrary linear combination of vectors ${\bf \Phi_{n}}$
(see next section for details) and
$|0\rangle$ is the pseudovacuum state, choosen
here as the standard ferromagnetic one
\begin{equation}
|0\rangle=\otimes _{j=1}^N|0\rangle_j,
\end{equation}
where
\begin{equation}
|0\rangle_i=\left(
\begin{array}{c}
1 \\
0
\end{array}
\right)_i\otimes \left(
\begin{array}{c}
1 \\
0
\end{array}
\right)_i
\end{equation}
which corresponds to the doubly occupied state.
Then it is easy to work out the action of
the monodromy matrices $T, T^{-1}$ on this state.
We begin by writing the periodic
monodromy matrix $T(u)$ as
\begin{equation}
T(u) =  \left( \begin {array} {cccc}
B(u)&B_1(u)&B_2(u)&F(u)\\
C_1(u)&A_{11}(u)&A_{12}(u)&E_1(u)\\
C_2(u)&A_{21}(u)&A_{22}(u)&E_2(u)\\
C_3(u)&C_4(u)&C_5(u)&D(u)
\end{array} \right).
\end{equation}
and $\bar{T}(u) =T^{-1}(-u)$ as
\begin{equation}
\bar{T}(u) =  \left( \begin {array} {cccc}
\bar{B}(u)&\bar{B}_1(u)&\bar{B}_2(u)&\bar{F}(u)\\
\bar{C}_1(u)&\bar{A}_{11}(u)&\bar{A}_{12}(u)&\bar{E}_1(u)\\
\bar{C}_2(u)&\bar{A}_{21}(u)&\bar{A}_{22}(u)&\bar{E}_2(u)\\
\bar{C}_3(u)&\bar{C}_4(u)&\bar{C}_5(u)&\bar{D}(u)\\
\end {array} \right).
\end{equation}
By computing the action of $T(u)$ and $\bar{T}(u)$ on the psedudovacuum
we find
\begin{eqnarray}
& &B(u) |0\rangle = \bar{B}(u)|0\rangle=|0\rangle,  \nonumber\\
& &D(u) |0\rangle = \bar{D}(u)|0\rangle=\left[u\right]^{2N}|0\rangle,  \nonumber\\
& &A_{aa}(u) |0\rangle = 
\bar{A}_{aa}(u)|0\rangle=\left[u\,\exp\left(\beta\right)\right]^{N}|0\rangle, \nonumber\\
& &A_{12}(u) |0\rangle = \bar{A}_{12}(u)|0\rangle=0,\nonumber\\
& &A_{21}(u) |0\rangle = \bar{A}_{21}(u)|0\rangle=0,\nonumber\\
& &B_a(u) |0\rangle  \neq 0,~~~
\bar{B}_a(u)|0\rangle \neq 0,\nonumber\\
& &E_a(u) |0\rangle \neq 0,~~~
\bar{E}_a(u)|0\rangle \neq 0,\nonumber\\
& &F(u) |0\rangle  \neq  0,~~~
\bar{F}(u)|0\rangle \neq 0,  \nonumber\\
& &C_i(u) |0\rangle =  \bar{C}_i(u)|0\rangle = 0,\nonumber\\
& &i=1,\cdots ,5,
a,b=1,2.\label{E-on-vacc}
\end{eqnarray}
Similarly, the doubled monodromy matrix
$U_{-}=T(u) K_{-}(u)\bar{T}(u)$(\ref{DMM}) can be arranged as
\begin{equation}
U_-(u) =   \left( \begin{array} {cccc}
\tilde{B}(u)&\tilde{B}_1(u)&\tilde{B}_2(u)&\tilde{F}(u)\\
\tilde{C}_1(u)&\tilde{A}_{11}(u)&\tilde{A}_{12}(u)&\tilde{E}_1(u)\\
\tilde{C}_2(u)&\tilde{A}_{21}(u)&\tilde{A}_{22}(u)&\tilde{E}_2(u)\\
\tilde{C}_3(u)&\tilde{C}_4(u)&\tilde{C}_5(u)&\tilde{D}(u)
\end{array}\right),\label{UM}
\end{equation}
Then using eq.(\ref{E-on-vacc}) and the Yang-Baxter algebra
\begin{equation}
\stackrel{2}{\bar{T}}(u)R_{12}(u,-u)\stackrel{1}{T}(u)=
\stackrel{1}{T}(u)R_{12}(u,-u)\stackrel{2}{\bar{T}}(u),\label{YBA}
\end{equation}
we can calculate the action of the doubled monodromy $U_{-}$
on the pseudovacuum, which due to the existence of two
classes of boundary $K_{-}$ matrices (\ref{Km1}), (\ref{Km2}) yields
(see Appendix B)
\begin{eqnarray}
& &\tilde{B}(u)|0\rangle = W_{1}^{-}(u)B(u)\bar{B}(u)|0\rangle, \label{Wm1-c}\\
& & \hat{A}_{aa}(u)|0\rangle  =
W^{-}_{a+1}(u)A_{aa}(u)\bar{A}_{aa}(u)|0\rangle, \label{fact-m}\\
& &\hat{D}(u)|0\rangle =W^{-}_{4}(u)D(u)\bar{D}(u)|0\rangle,\label{Wm4-c} \\
& &\tilde{B}_a(u) |0\rangle \neq  0,~~~
\tilde{E}_a(u) |0\rangle \neq 0, \nonumber\\
& &\tilde{A}_{ab}(u) |0\rangle = 0,~~~
\tilde{F}(u) \neq 0, \nonumber\\
& &\tilde{C}_i(u) |0\rangle  =  0 ,~~~
i=1,\cdots ,5,\,\,a\neq b=1,2,\label{Wmc-c}
\end{eqnarray}
where for case I:
\begin{eqnarray}
& &W^{-}_{1}(u)=1,\nonumber\\
& &W^{-}_{2}(u)=W^{-}_{3}(u)=-\frac{2h^2u\left(uc_1-1\right)}{\left(h^2u^2-1\right)\left(h^2u-c_1\right)},\nonumber\\
& &W^{-}_{4}(u)=\frac{2u^2\left(uc_1-1\right)\left(h^2+1\right)\left(uc_1-h^2\right)}{\left(u^2-1\right)\left(u^2-h^2\right)\left(h^2u^2-c_1\right)\left(u-c_1\right)},\nonumber
\end{eqnarray}
and for case II:
\begin{eqnarray}
& &W^{-}_{1}(u)=1,\nonumber\\
& &W^{-}_{2}(u)=\frac{2h^2u^2}{\left(h^2u^2-1\right)},\nonumber\\
& 
&W^{-}_{3}(u)=\frac{2hu\left(hu+c_1\right)}{\left(h^2u^2-1\right)\left(1+c_1hu\right)},\nonumber
\\
& 
&W^{-}_{4}(u)=\frac{2u^3\left(h^2+1\right)\left(u+hc_1\right)}{\left(u^2-1\right)\left(u^2-h^2\right)\left(1+c_1hu\right)},\nonumber
\end{eqnarray}
Actually, in order to considerably simplify the calculations, the following
transformations
\begin{eqnarray}
\hat{A}_{aa}(u) & = & \tilde{A}_{aa}(u)-\frac{
\rho_2(u,-u)}{ \rho _1(u,-u)}\tilde{B}(u), \nonumber\\
\hat{D}(u) & = & \tilde{D}(u)-\frac{\rho _4(u,-u)}{\rho
_1(u,-u)}\tilde{B}(u)\nonumber\\
& &- \frac{1+u^2}{h^2-u^2}\hat{A}
_{11}(u)-\frac{h^2(1+u^2)}{h^2-u^2}\hat{A}
_{22}(u),\label{trans}
\end{eqnarray}
were used in the above equations (\ref{Wm1-c})--(\ref{Wm4-c}).
In fact, these types of transformations have also been performed
in other models with OBC \cite{OPC4,OPC1,OPC2,OPC3}. They simplify not only
these expressions for the action of the operators on the pseudovacuum,
but also simplify the fundamental commutation relations,
allowing for an easy recognition of the
``wanted'' and ``unwanted'' terms, essential to
the Bethe ansatz machinery (see next section for details).
Then, after this transformation and taking into account the
two sets of boundary matrices $K_{+}$ (\ref{Kp1}),(\ref{Kp2}), 
the transfer matrix (\ref{DTM})
can be recast as
\begin{equation}
\tau(u) =
 W^+_1(u)\tilde{B}(u)+\sum _{a=1}^2 W^+_{a+1}(u)\hat{A}_{aa}(u)+W^+_4(u)\hat{D}(u),  
\label{DTM-E}
\end{equation}
where for case I:
\begin{eqnarray}
& 
&W^{+}_{1}(u)=\frac{2u^2\left(u-c_2\right)\left(h^2+1\right)\left(h^2u-c_2\right)}{\left(uc_2+1\right)\left(uh^2c_2+1\right)\left(u^2-1\right)\left(h^2u^2-1\right)},\nonumber\\
& &W^{+}_{2}(u)=\frac{u\left(uc_2-1\right)\left(u-c_2\right)\left(h^2+1\right)}{\left(h^2uc_2+1\right)\left(u^2-h^2\right)\left(uc_2+1\right)},\nonumber\\
& &W^{+}_{3}(u)=\frac{h^2u\left(uc_2-1\right)\left(u-c_2\right)\left(h^2+1\right)}{\left(h^2uc_2+1\right)\left(u^2-h^2\right)\left(uc_2+1\right)},\nonumber\\
& &W^{+}_{4}(u)=\frac{\left(uc_2-1\right)\left(uc_2-h^2\right)}{\left(uc_2+1\right)\left(h^2uc_2+1\right)},\nonumber
\end{eqnarray}
and for case II:
\begin{eqnarray}
& &W^{+}_{1}(u)=\frac{2u\left(h^2+1\right)\left(hc_2u+1\right)}{h\left(u^2-1\right)\left(h^2u^2-1\right)\left(hu-c_2\right)},\nonumber\\
& &W^{+}_{2}(u)=-\frac{\left(h^2+1\right)\left(u+hc_2\right)}{h\left(u^2-h^2\right)\left(hu-c_2\right)}\nonumber\\
& &W^{+}_{3}(u)=-\frac{u\left(h^2+1\right)\left(h+uc_2\right)}{\left(u^2-h^2\right)\left(hu-c_2\right)}\nonumber\\
& &W^{+}_{4}(u)=\frac{\left(u+hc_2\right)}{h\left(hu-c_2\right)},\nonumber
\end{eqnarray}
Notice from eqs. (\ref{Wm1-c})--(\ref{Wmc-c}) that the Lax operators (\ref{UM})  act as
creation (annihilation) operators on the pseudovacuum.
Before ending this section, we would like to recall that the boundary
matrix pairs $\left(K_{+}(u),K_{-}(u)\right)$ lead to four
kinds of boundary fields preserving the integrability of the
model. Consequently, the factors $W^{\pm }_i(k)$ $i=1,\ldots ,4$ also form a
pair $(W^{+}_i(k),W^{-}_i(k))$ leading to four possible choices consistent
with the boundary matrix pairs.  The perfect factorization
provide us with the nesting boundary $K$-matices appearing in
(\ref{fact-m}) and (\ref{DTM-E}) respectively. In the following
section, we shall prove that they indeed constitute the solutions of the RE
for the hidden asymmetric $XXZ$ spin open chain.

\section{Algebraic Bethe ansatz II: The nesting procedure}
\label{sec4}
The next step in the solution of the eigenvalue problem of the transfer
matrix consists in the computation of the
necessary commutation relations
between the diagonal fields and the creation fields. From the
RE (\ref{RE1}) and the definition (\ref{UM}), after tedious calculations,
we obtain the main fundamental commutation relations. We display below just those needed for 
future calculations
\begin{eqnarray}
& &\tilde{B}(u)\tilde{B}_a(v) = \frac{(u+v)(uvh^2+1)}{(u-v)(uvh^2-1)} \tilde{B}_a(v) 
\tilde{B}(u)+u.t.,  \label{commu1}\\
& &\hat{D}(u)\tilde{B}_a(v) = 
\frac{(u+v)(uvh^2+1)}{(u-vh^2)(uv-1)}\tilde{B}_a(v)\hat{D}(u)+u.t.,  \label{commu2}\\
& & \hat{A}_{ad_1}(u)\tilde{B}_{d_2}(v) = 
\frac{(u+v)(1+uvh^2)}{(u-v)(uvh^2-1)}\bar{r}_{12}(u,-v)^{c_1b_2}_{a_1c_2}\bar{r}_{21}(-v,-u)^{d_
1d_2}_{b_1b_2}\nonumber\\
& &\times \tilde{B}_{c_2}(v)\hat{A}_{c_1b_1}(u)+u.t.,\label{commu3}\\
& & \tilde{B}_a(u)\otimes \tilde{B}_b(v) = \left\{\tilde{B}_c(v)\otimes
\tilde{B}_d(u)\right.\nonumber\\
& &
\left.+\frac{\sqrt{1+h^2u^2}\sqrt{1+v^2}}{1-uv}\tilde{F}(v)\vec{\xi}
\left(I\otimes \tilde{A}(u)\right)\right\}.r(-v,-u)  \nonumber\\
& &-\frac{\sqrt{1+h^2v^2}\sqrt{1+u^2}}{1-uv}\tilde{F}(u)\vec{\xi }\left(I\otimes 
\tilde{A}(v)\right)  \nonumber\\
& &
+\frac{\left(h^2u+v\right)\sqrt{1+h^2u^2}\sqrt{1+v^2}}{\left(u-h^2v\right)\left(uv-1\right)}\left[\tilde{F}(v)\tilde{B}(u)\right.\nonumber\\
& &
\left.-\frac{\left(h^2v+u\right)\sqrt{1+h^2v^2}\sqrt{1+u^2}}{\left(h^2u+v\right)\sqrt{1+h^2u^2}\sqrt{1+v^2}}\tilde{F}(u)\tilde{B}(v)\right].\vec{\xi},
\label{commu4}
\end{eqnarray}
where
\begin{eqnarray}
& & \vec{\xi }=(0,h^{-1},1,0), \,\,\tilde{A}(u)=\left( \begin {array} {cc}
\tilde{A}_{11}(u)&\tilde{A}_{12}(u)\\
\tilde{A}_{21}(u)&\tilde{A}_{22}(u)
\end {array} \right),\\
& &
{\bar{r}_{12}}(u,-v) = \left( \begin {array} {cccc}
1&0&0&0\\
0&\bar{c}(u,-v)&\bar{d}(u,-v)&0\\
0&\bar{a}(u,-v)&\bar{c}(u,-v)&0\\
0&0&0&1
\end {array} \right),  \\
& &r(-v,-u) = \left( \begin {array} {cccc}
1&0&0&0\\
0&a(-v,-u)&c(-v,-u)&0\\
0&c(-v,-u)&d(-v,-u)&0\\
0&0&0&1
\end {array} \right),
\end{eqnarray}
and
\begin{eqnarray}
& &\bar{a}(u,-v) = \frac{\rho _7(u,-v)\rho _1(u,-v)-\rho_2^2(u,-v)}{\rho _1^2(u,-v)-\rho
_{2}^2(u,-v)},\nonumber\\
& &\bar{d}(u,-v)=\frac{\rho _{10}(u,-v)\rho _1(u,-v)-\rho_2^2(u,-v)}{\rho 
_1^2(u,-v)-\rho_{2}^2(u,-v)}\nonumber\\
& &\bar{c}(u,-v)=\frac{\rho _{1}(u,-v)\rho _{15}(u,-v)}{\rho _1^2(u,-v)-\rho
_{2}^2(u,-v)}\nonumber\\
& &a(-v,-u) = \frac{\rho _{14}(-v,-u)\rho _{10}(-v,-u)-\rho
_{13}^2(-v,-u)}{\rho _1(-v,-u)\rho _{14}(-v,-u)},\nonumber\\
& &d(-v,-u)=\frac{\rho _{14}(-v,-u)\rho _{7}(-v,-u)-\rho _{12}^2(-v,-u)}{\rho _1(-v,-u)\rho 
_{14}(-v,-u)},\nonumber\\
& &
c(-v,-u) =\frac{\rho _{14}(-v,-u)\rho _{15}(-v,-u)-\rho_{12}(-v,-u)\rho
_{13}(-v,-u)}{\rho _1(-v,-u)\rho _{14}(-v,-u)}.
\end{eqnarray}
It remains to consider the commutation relations for the creation
field $\tilde{F}$. Since they are extremely involved, we have
collected them in Appendix B.  We remark here that the fundamental
commutation relations are much more complicated in comparsion with
those obtained in the case of periodic boundary conditions, turning
the calculations here much more cumbersome and difficult to treat.  In
all  relations, we have omitted the
unwanted terms (those that can not generate an eigenvector of the
transfer matrix) to save space. It turns out that the auxiliary
matrices $\bar{r}_{12}(u,-v)$ and $r(-v,-u)$ are nothing but the
rational $R$ matrices of an isotropic asymmetric six-vertex model.
Indeed, by performing the parameterization
$u=\exp\left(\mathrm{i}k_1\right), v=\exp\left(\mathrm{i}k_2\right)$ and
$h=\exp\left(\beta\right)=\exp\left(-\mathrm{i}\eta\right)$, one may
present these matrices as
\begin{eqnarray}
& &r(-v,-u)=r(k_1-k_2)=\left( \begin {array} {cccc}
1&0&0&0\\
0&a(k_1-k_2)&c(k_1-k_2)&0\\
0&c(k_1-k_2)&d(k_1-k_2)&0\\
0&0&0&1
\end {array} \right), \label{nest-r} \\
& &\bar{r}_{12}(u,-v) =P.r(k_1+k_2-2\eta ),\\
& &\bar{r}_{21}(u,-v) =P.r(k_1-k_2).P
\end{eqnarray}
with the Boltzmann weights
\begin{eqnarray}
 & &
a(k)=\frac{\sin\eta 
}{\sin\left(\frac{k}{2}+\eta\right)}\exp\left(-\mathrm{i}\frac{1}{2}k\right),\\
& &
d(k)=\frac{\sin\eta 
}{\sin\left(\frac{k}{2}+\eta\right)}\exp\left(\mathrm{i}\frac{1}{2}k\right),\\
& &
c(k)=-\frac{\sin\frac{k}{2} 
}{\sin\left(\frac{k}{2}+\eta\right)}\exp\left(-\mathrm{i}\frac{1}{2}k\right).
\end{eqnarray}
In view of the commutation relation (\ref{commu4}), we may phenomenologically write \cite{MR}
the $n$-particle state as
\begin{equation}
|\Phi_n(k_1,\ldots,k_n)\rangle=
\Phi_{n}(k_1,\ldots,k_n){\bf F}|0\rangle .
\label{np-state1}
\end{equation}
with a recursive relation of $n$-particle vector
\begin{eqnarray}
& & 
\Phi _{n}(k_1,\ldots,k_n)
= \tilde{B}_{a_1}(k_1)\otimes\Phi_{n-1}(k_2,\ldots,k_n)  \nonumber\\
& & +\sum _{j=2}^{n}\left\{\left[\vec{\xi}\otimes \tilde{F}(k_1)\right] 
 \Phi_{n-2}(k_2,\ldots,k_{j-1},k_{j+1},\ldots,k_n)\right.\nonumber\\
& & \hphantom{+\sum _{j=2}^{n}}\left.\times 
 \tilde{B}(k_j)G^{(n)}_{j-1}(k_1,\ldots,k_n)\right\}\nonumber\\
& &+\sum _{j=2}^{n}\left\{\left[\vec{\xi}\otimes \tilde{F}(k_1)\right] 
 \Phi_{n-2}(k_2,\ldots,k_{j-1},k_{j+1},\ldots,k_n)\right.\nonumber\\
& & \hphantom{-\sum _{j=2}^{n}}\left.\times 
 \left[I\otimes \tilde{A}(k_j)\right]H^{(n)}_{j-1}(k_1,\ldots,k_n)\right\}.
 \label{np-vector}
\end{eqnarray}
Above  ${\bf F}=F^{\mbox{\scriptsize $a_1,\ldots ,a_n$}}$ are the coefficients
of arbitrary linear combination of the vectors reflecting the `spin'
degrees of freedom with $a_i=1,2$ ($i=1,\ldots,n$) and $\vec{\xi}$
plays the role of forbidding two spin up or two spin down particles at
the same site. $\tilde{F}(u)$ creates a local particle pair with opposite
spins. From the commutation relation (\ref{commu4}), we can also conclude that $\Phi
_{n}(k_1,\cdots,k_n)$ satisfies the symmetry relation 
\begin{eqnarray}
& &\Phi _{n}(k_1,\cdots,k_j,k_{j+1},\cdots,k_n) = \nonumber\\
& &\Phi _{n}(k_1,\cdots,k_{j+1},k_{j},\cdots,k_n).r(k_{j}-k_{j+1})
\end{eqnarray}
This symmetry giving a restriction to the functions
$H^{(n)}_{j-1}(k_1, \cdots,k_n)$ and $G^{(n)}_{j-1}(k_1,\cdots,k_n)$
is very useful to deduce the coefficients and simplify the unwanted
terms in the eigenvalue of the transfer matrix. In fact, after
performing three-particle scattering, the explicit form of these
coefficients can be fixed. However checking three-particle scattering is
indeed a extremely tough calculation and we had to leave the coefficients to be
determined later. Explicitly, we display the two-particle state
\begin{eqnarray}
& &\Phi _{2}(k_1,k_2) =  \tilde{B}_{\mbox{\scriptsize $a_1$}}(k_1)\otimes 
\tilde{B}_{\mbox{\scriptsize $a_2$}}(k_2)\nonumber \\
& 
&+\frac{\sqrt{1+e^{\mathrm{i}2k_1}}\sqrt{1+e^{\mathrm{i}2(k_2-\eta)}}}{1-e^{\mathrm{i}(k_1+k_2)}
}\tilde{F}(k_1)\stackrel{\rightarrow}{\xi }(I\otimes \tilde{A}(k_2)) \nonumber \\
& &+\frac{\cos(\frac{k_1-k_2}{2}+\eta 
)\sqrt{1+e^{\mathrm{i}2k_1}}\sqrt{1+e^{\mathrm{i}2(k_2-\eta)}}}{\sin(\frac{k_1-k_2}{2}+\eta 
)(1-e^{\mathrm{i}2(k_1+k_2)})}\tilde{F}(k_1)\tilde{B}(k_2).\stackrel{\rightarrow}{\xi }.
\end{eqnarray}
Implementing a rescaling $k\rightarrow k+\eta$ and seting 
$c_1=e^{\mathrm{i}2\xi_-},c_2=e^{\mathrm{i}2\xi_+}$, the eigenvalue of the transfer matrix
(\ref{DTM}) acting on the state (\ref{np-state1}) can be given as 
\begin{eqnarray}
& &\tau(k)|\Phi_n(k_1,\ldots,k_n)\rangle=\left\{W^{+}_{1}(k)W^{-}_{1}(k)
\prod_{i=1}^n\frac{\cos\left(\frac{k+k_i}{2}\right)\cos\left(\frac{k-k_i}{2}\right)}{\sin\left(\frac{k+k_i}{2}\right)\sin\left(\frac{k-k_i}{2}\right)}\right.\nonumber\\
& &
\left.+\sum_{a=1}^{2}{W}^{+}_{a+1}(k){W}^{-}_{a+1}(k)e^{\mathrm{i}2Nk}\right.\nonumber\\
& &
\left.\times 
\prod_{i=1}^n\frac{\cos\left(\frac{k+k_i}{2}\right)\cos\left(\frac{k-k_i}{2}\right)}{\sin\left(\frac{k+k_i}{2}\right)\sin\left(\frac{k-k_i}{2}\right)}\Lambda^{(1)}_{aa}\left(k,\{k_i\},\{\lambda
a_j\}\right)\right.\nonumber\\
& &
\left.+W^{+}_{4}(k)W^{-}_{4}(k)e^{\mathrm{i}4N(k+\eta)}
\prod_{i=1}^n\frac{\cos\left(\frac{k+k_i}{2}\right)\cos\left(\frac{k-k_i}{2}\right)}{\sin\left(\frac{k+k_i}{2}+\eta\right)\sin\left(\frac{k-k_i}{2}+\eta\right)}\right\}\nonumber\\
& &|\Phi_n(k_1,\ldots,k_n)\rangle
\end{eqnarray}
provided that 
\begin{eqnarray}
\frac{W^+_1(k)W^{-}_1(k)}{W^+_2(k)W^{-}_{2}(k)exp\left(\mathrm{i}2Nk\right)}
\mid _{\mbox{\scriptsize $k=k_i$}}=-\Lambda ^{(1)}\left(k=k_i,\{k_i\},\{\lambda _j\}\right), 
\label{bethe-1}\\
& &i=1,\cdots n.\nonumber
\end{eqnarray}
Here $\Lambda ^{(1)}\left(\tilde{k},\{k_i\},\{\lambda _j\}\right)$ is
the eigenvalue of the reduced transfer matrix $\tau ^{(1)}(k)$, which  turns out to be
an inhomogenous isotropic six-vertex model with open boundary conditions, namely,
\begin{equation}
\tau^{(1)}(k)F^{\mbox{\scriptsize $a_1\cdots a_n$}
}=\Lambda ^{(1)}\left(k,\{k_i\},\{\lambda _j\}\right)F^{\mbox{\scriptsize $a_1\cdots
a_n$}},  \label{nest-e}
\end{equation}
where 
\begin{equation}
\tau^{(1)}(k)={\rm Tr}_0K^{(1)}_{+}(k)T^{(1)}(k)K^{(1)}_{-}(k){T^{(1)}}
^{-1}(-k).  \label{nest-TM}
\end{equation}
The nesting monodromy matrices $T^{(1)}(k)$ and ${T^{(1)}}^{-1}(-
k)$ read 
\begin{eqnarray}
T^{(1)}(k) & =  & r_{12}(k+k_1)_{\mbox{\scriptsize
$ac_1$}}^{\mbox{\scriptsize $a_1e_1$}}
r_{12}(k+k_2)_{\mbox{\scriptsize
$a_1c_2$}}^{\mbox{\scriptsize $a_2e_2$}}
\cdots r_{12}(k+k_n)_{
\mbox{\scriptsize $a_{n-1}c_n$}}^{\mbox{\scriptsize $a_ne_n$}}, \\
{T^{(1)}}^{-1}(-k) &  =  & r_{21}(k-k_n)_{
\mbox{\scriptsize $b_ne_n$}}^{\mbox{\scriptsize $b_{n-1}d_n$}}\cdots r_{21}(
k-k_2)_{\mbox{\scriptsize $b_2e_2$}}^{\mbox{\scriptsize
$b_1d_2$}}r_{21}(k-k_1)_{\mbox{\scriptsize $b_1e_1$}}^{\mbox{\scriptsize
$ad_1$}},
\end{eqnarray}
and the nesting $K^{(1)}_{\pm}(k)$ may be chosen as\\
Case I
\begin{equation}
K^{(1)}_{+}(k)=\left( \begin{array}{cc}
e^{\mathrm{i}\eta}&0\\
0&e^{-\mathrm{i}\eta}
\end{array} \right),
\label{nest-K-1}
\end{equation}
in such a way that the factors $W^{+}$ can be rewritten as
\begin{eqnarray}
& &W^{+}_{1}(k)=\frac{\sin(\frac{k}{2}+\frac{\eta }{2}-\xi _+)
\sin(\frac{k}{2}-\frac{\eta }{2}-\xi _+)\cos\eta}
{\cos(\frac{k}{2}+\frac{\eta }{2}+\xi _+)\cos(\frac{k}{2}-\frac{\eta }{2}+\xi _+)\sin 
k\sin(k+\eta)},\label{Wp1-1}\\
& &
W^{+}_{2}(k)=W^{+}_{3}(k)=\frac{\sin(\frac{k}{2}+\frac{\eta }{2}+\xi _+)
\sin(\frac{k}{2}+\frac{\eta }{2}-\xi _+)\cos\eta}
{\cos(\frac{k}{2}-\frac{\eta }{2}+\xi _+)\cos(\frac{k}{2}+\frac{\eta }{2}+\xi 
_+)\sin(k+2\eta)},\\
& &
W^{+}_{4}(k)=\frac{\sin(\frac{k}{2}+\frac{\eta }{2}+\xi _+)\sin(\frac{k}{2}+\frac{3\eta }{2}+\xi 
_+)}
{\cos(\frac{k}{2}+\frac{\eta }{2}+\xi _+)\cos(\frac{k}{2}-\frac{\eta }{2}+\xi _+)},
\end{eqnarray}
and 
\begin{equation}
K^{(1)}_{-}(k)=\left( \begin{array} {cc}
1&0\\
0&1
\end{array}\right),\label{nest-K-2}
\end{equation}
with
\begin{eqnarray}
& &W^{-}_{1}(k)=1,\\
& &
W^{-}_{2}(k)=W^{-}_{3}(k)=-\frac{\sin(\frac{k}{2}+\frac{\eta }{2}+\xi 
_-)}{\sin(\frac{k}{2}-\frac{\eta }{2}-\xi _-)\sin k },\\
& &
W^{-}_{4}(k)=\frac{\sin(\frac{k}{2}+\frac{\eta }{2}+\xi _-)\sin(\frac{k}{2}+\frac{3\eta }{2}+\xi 
_-)\cos\eta}
{\sin(\frac{k}{2}-\frac{\eta }{2}-\xi _-)\sin(\frac{k}{2}+\frac{\eta }{2}-\xi_-)\sin 
(k+\eta)\sin(k+2\eta)},
\end{eqnarray}
Case II
\begin{equation}
K^{(1)}_{+}(k)=\left(\begin{array}{cc}
\cos(\frac{k}{2}+\eta -\xi_+)e^{-\mathrm{i}k}&0\\
0&\cos(\frac{k}{2}+\eta+\xi _+)
\end{array} \right),
\label{nest-K-3}
\end{equation}
with
\begin{eqnarray}
& &W^{+}_{1}(k)=\frac{\cos(\frac{k}{2}+\xi _+)\cos \eta e^{-\mathrm{i}k}}
{\sin(\frac{k}{2}-\xi _+)\sin k\sin(k+\eta)},\\
& &
W^{+}_{2}(k)=W^{+}_{3}(k)=-\frac{\cos\eta}{\sin(\frac{k}{2}-\xi  _+)\sin(k+2\eta)},\\
& &
W^{+}_{4}(k)=\frac{\cos(\frac{k}{2}+\eta-\xi _+)e^{\mathrm{i}\eta }}{\sin(\frac{k}{2}-\xi _+)},
\end{eqnarray}
and
\begin{equation}
K^{(1)}_{-}(k)=\left( \begin{array}{cc}
\cos(\frac{k}{2}+\xi_-)e^{\mathrm{i}k}&0\\
0&\cos(\frac{k}{2}-\xi_-)
\end{array} \right),\label{nest-K-4}
\end{equation}
with 
\begin{eqnarray}
& &W^{-}_{1}(k)=1,\\
& &
W^{-}_{2}(k)=W^{-}_{3}(k)=\frac{1}{\sin k\cos(\frac{k}{2}+\xi _-)},\\
& &
W^{-}_{4}(k)=\frac{\cos(\frac{k}{2}+\eta -\xi _-)\cos\eta e^{\mathrm{i}(k+\eta )} }
{\cos(\frac{k}{2}+\xi _-)\sin (k+\eta)\sin(k+2\eta)}.\label{Wm2-2}
\end{eqnarray}
We notice that the nesting boundary $K^{(1)}_{\pm }(k)$-matrices
also constitute four classes of BC consistent with the boundary
matrix pairs $(K^{+}(k),K^{-}(k))$.
So far, the eigenvalue problem of the 1D Bariev model with four kinds of possible boundary 
fields 
reduces to the solution of the nested auxiliary transfer matrix (\ref{nest-e}), which
can be associated with an inhomogeneous isotropic six-vertex model with open boundary conditions 
(\ref{nest-K-1})--(\ref{nest-K-4}). Furthermore, we can 
check that the $r$-matrix (\ref{nest-r}) and the nesting $K^{(1)}_{\pm}$-matrices 
(\ref{nest-K-1})-- (\ref{nest-K-4}),
which arises from the factorization of the spin degree of freedom,
indeed  satisfy the Yang-Baxter algebra 
\begin{equation}
r_{12}(k-\lambda )\stackrel{1}{T^{(1)}}(k)\stackrel{2}{T^{(1)}}(\lambda )
=\stackrel{2}{T^{(1)}}(\lambda )\stackrel{1}{T^{(1)}}(
k)r_{12}(k-\lambda ),
\label{nest-ybe}
\end{equation}
and the reflection equations 
\begin{eqnarray}
& &r_{12}(k-\lambda )\stackrel{1}{K_{-}^{(1)}}(k)r_{21}(k+\lambda 
)\stackrel{2}{K_{-}^{(1)}}(\lambda )\nonumber\\
& &
=\stackrel{2}{K_{-}^{(1)}}(\lambda 
)r_{12}(k+\lambda)\stackrel{1}{K_{-}^{(1)}}(k)r_{21}(k-\lambda),\label{nest-re-1}\\
& &r_{21}^{{\rm t}_1{\rm t}_2}(\lambda-k)\stackrel{1}{K_+^{(1)
~{\rm t}_1}}(k)\tilde{r}_{12}(-k-\lambda ) 
\stackrel{2}{K_+^{(1)~{\rm t}_2}}(\lambda) \nonumber \\
& &=\stackrel{2}{K_+^{(1)~{\rm t}_2}}(\lambda )\tilde{r}_{21}(-k-\lambda )
\stackrel{1}{K_+^{(1)~{\rm t}_1}}(k) r
_{12}^{{\rm t}_1{\rm t}_2}(\lambda-k),  \label{nest-re-2}
\end{eqnarray}
where 
\begin{eqnarray}
r_{12}(k)r_{21}(-k) & = & 1,\\
\tilde{r}_{21}^{{\rm t}_1}(-k)r_{12}^{{\rm t}_2}(k) & = & 1,\\
\tilde{r}_{12}^{{\rm t}_2}(-k)r_{21}^{{\rm t}_1}(k) & = & 1.
\end{eqnarray}
Following all steps in solving $XXZ$ open chain in \cite{EK}, one can completely
diagonalize the nesting transfer matrix (\ref{nest-TM}) with the help of 
the main ingredients (\ref{nest-ybe})
-(\ref{nest-re-2}) describing the open BC for the hidden
asymmetric $XXZ$ open chain.
Therefore, what follows is the exact solution of the $XXZ$ open
chain characterizing the spin degree of freedom, i.e., 
\begin{eqnarray}
& &\Lambda ^{(1)}(k,\{k_i\},\{\lambda _j\})\mid \Phi ^{(1)}(\lambda_1,\cdots,\lambda _M)\rangle 
\nonumber\\
& &=
\left\{\frac{\sin(k+2\eta)}{\sin(k+\eta)}W^{+}_{11}(k)W^{-}_{11}(k)A^{(1)}(k)
\prod_{j=1}^{M}\frac{\sin(\frac{k+\lambda_j}{2})\sin(\frac{k-\lambda_j}{2}-\eta)}
{\sin(\frac{k+\lambda_j}{2}+\eta)\sin(\frac{k-\lambda_j}{2})}\right.\nonumber\\
& &
\left.+\frac{\sin k}{\sin(k+\eta)}W^{+}_{22}(k)W^{-}_{22}(k)D^{(1)}(k)
\prod_{j=1}^{M}\frac{\sin(\frac{k+\lambda_j}{2}+2\eta)\sin(\frac{k-\lambda_j}{2}+\eta)}
{\sin(\frac{k+\lambda_j}{2}+\eta)\sin(\frac{k-\lambda_j}{2})}\right\}\nonumber\\
& &
\times\mid \Phi ^{(1)}(\lambda_l,\cdots,\lambda _M)\rangle,
\end{eqnarray}
provided that 
\begin{equation}
\frac{W^{+}_{11}(k)W^{-}_{11}(k)A^{(1)}(k)}
{W^{+}_{22}(k)W^{-}_{22}(k)D^{(1)}(k)}
=\prod_{
\begin{array}{l}
l=1, \\ 
l\neq j
\end{array}}^{M}
\frac{\sin(\frac{\lambda_j+\lambda_l}{2}+2\eta)\sin(\frac{\lambda_j-\lambda_l}{2}+\eta)}
{\sin(\frac{\lambda_j+\lambda_l}{2})\sin(\frac{\lambda_j-\lambda_l}{2}-\eta)},
\end{equation}
where 
\begin{eqnarray}
& &A^{(1)}(k)\mid \Phi ^{(1)}\rangle=\mid \Phi ^{(1)}\rangle,\\
& &
D^{(1)}(k)\mid \Phi 
^{(1)}\rangle=\left\{\prod_{i=1}^{n}\frac{\sin(\frac{k+\lambda_i}{2})\sin(\frac{k-\lambda_i}{2})
}
{\sin(\frac{k+\lambda_i}{2}+\eta)\sin(\frac{k-\lambda_i}{2}+
\eta)}\right\}\mid \Phi ^{(1)}\rangle.
\end{eqnarray}
Here for  case I, we have
\begin{eqnarray}
& &W^{+}_{11}(k)=W^{-}_{11}(k)=1,\\
& &
W^{+}_{22}(k)=W^{-}_{22}(k)=1,
\end{eqnarray}
while for case II,
\begin{eqnarray}
& 
&W^{+}_{11}(k)=\cos(\frac{k}{2}-\xi_+)e^{-\mathrm{i}k},\,\,W^{-}_{11}(k)=\cos(\frac{k}{2}+\xi_-)
e^{\mathrm{i}k},\\
& &
W^{+}_{22}(k)=\cos(\frac{k}{2}+\eta+\xi_+)e^{\mathrm{i}\eta},\,\,W^{-}_{22}(k)=\cos(\frac{k}{2}+
\eta-\xi_-)e^{-\mathrm{i}\eta},
\end{eqnarray}
respectively. 

Finally, making a shift on the spin rapidity $\lambda
_i\rightarrow \lambda _i-\eta$, the eigenvalue of the transfer matrix (\ref{DTM}) with two 
classes of boundary $K$-matrices can be given explicitly
(up to a common factor)
\begin{eqnarray}
& &
\tau(u)\mid \Phi _n(k_1,\cdots,k_n)\rangle =\left\{
\prod_{i=1}^{n}\frac{\cos(\frac{k+k_i}{2})\cos(\frac{k-k_i}{2})}
{\sin(\frac{k+k_i}{2})\sin(\frac{k-k_i}{2})}\left\{W^{+}_{1}(k)W^{-}_1(k)\right.\right.\nonumber
\\
& &
\left.\left.+\frac{\sin(k+2\eta)}{\sin(k+\eta)}W^{+}_{2}(k)W^{-}_2(k)W^{+}_{11}(k)W^{-}_{11}(k)e
^{\mathrm{i}2Nk}\right.\right.\nonumber\\
& &
\left.\left.\times \prod_{l=1}^{M}
\frac{\sin(\frac{k+\lambda_l-\eta}{2})\sin(\frac{k-\lambda_l-\eta}{2})}
{\sin(\frac{k+\lambda_l+\eta}{2})\sin(\frac{k-\lambda_l+\eta}{2})}\right\}\right.\nonumber\\
& &
\left.+\prod_{i=1}^{n}\frac{\cos(\frac{k+k_i}{2})\cos(\frac{k-k_i}{2})}
{\sin(\frac{k+k_i}{2}+\eta)\sin(\frac{k-k_i}{2}+\eta)}\left\{W^{+}_{4}(k)W^{-}_4(k)e^{\mathrm{i}
4N(k+\eta)}\right.\right.\nonumber\\
& &
\left.\left.+\frac{\sin 
k}{\sin(k+\eta)}W^{+}_{3}(k)W^{-}_3(k)W^{+}_{22}(k)W^{-}_{22}(k)e^{\mathrm{i}2Nk}\right.\right.\nonumber\\
& &
\left.\left.\times\prod_{l=1}^{M}
\frac{\sin(\frac{k+\lambda_l+3\eta}{2})\sin(\frac{k-\lambda_l+3\eta}{2})}
{\sin(\frac{k+\lambda_l+\eta}{2})\sin(\frac{k-\lambda_l+\eta}{2})}\right\}\right\}
\mid \Phi _n(k_1,\cdots,k_n)\rangle \label{eigen},
\end{eqnarray}
where all factorized coefficients $W^{\pm}_i$ ($i=1,\cdots,4$, $11$
and $22$) have been given explicitly in the above equations, which provide four kinds of 
possible exact solutions. 
Finally, the expression above is in fact the eigenvalue of the
transfer matrix, if the following Bethe ansatz equations are satisfied
for the charge and spin rapidities $k_i, \, \lambda _j$ .
\begin{eqnarray}
e^{\mathrm{i}2Nk_i}\zeta_1(k_{i},\xi_{+}) \zeta_2(k_{i},\xi_{-})=
\prod_{l=1}^{M}
\frac{\sin(\frac{k_i+\lambda_l+\eta}{2})\sin(\frac{k_i-\lambda_l+\eta}{2})}
{\sin(\frac{k_i+\lambda_l-\eta}{2})\sin(\frac{k_i-\lambda_l-\eta}{2})},\label{Bethe1}\\
\psi _1(\lambda _j,\xi_{+}) \psi_2(\lambda _j,\xi_{-})\prod_{l=1}^{n}
 \frac{\sin(\frac{\lambda_j+k_l+\eta}{2})\sin(\frac{\lambda_j-k_l+\eta}{2})}
{\sin(\frac{\lambda_j+k_l-\eta}{2})\sin(\frac{\lambda_j-k_l-\eta}{2})}\nonumber\\
=\prod_{
\begin{array}{l}
l=1, \\ 
l\neq j
\end{array}}^{M}
\frac{\sin(\frac{\lambda_j+\lambda_l}{2}+\eta)\sin(\frac{\lambda_j-\lambda_l}{2}+\eta)}
{\sin(\frac{\lambda_j+\lambda_l}{2}-\eta)\sin(\frac{\lambda_j-\lambda_l}{2}-\eta)},\label{Bethe2}\\
j=1,\ldots ,M,\,\,i=1,\ldots ,n.\nonumber
\end{eqnarray}
The expressions for $\zeta_a(k_{i},\xi_{\pm })$ and $\psi _a(\lambda
_j,\xi_{\pm}),\, a=1,2$  depend on which type of boundaries we are
considering. Below we display all cases \\ Case (I,I)
\begin{eqnarray}
& &\zeta_1(k_{i},\xi_{+})=\frac{\sin(\frac{k_i+\eta}{2}+\xi _+)}
{\sin(\frac{k_i-\eta}{2}-\xi _+)},\,\,
\zeta_2(k_i,\xi_{-})=\frac{\sin(\frac{k_i+\eta}{2}+\xi _-)}
{\sin(\frac{k_i-\eta}{2}-\xi _-)},\\
& &
\psi _1(\lambda _j,\xi_{+})= \psi_2(\lambda _j,\xi_{-})=1,
\end{eqnarray}
Case (II,II)
\begin{eqnarray}
& &\zeta_1(k_{i},\xi_{+})=-\frac{\cos(\frac{k_i}{2}-\xi _+)}
{\cos(\frac{k_i}{2}+\xi _+)},\,\,
\zeta_2(k_i,\xi_{-})=-e^{\mathrm{i}k_i},\\
& &
\psi _1(\lambda _j,\xi_{+})=\frac{\cos(\frac{\lambda_j-\eta}{2}-\xi _+)e^{-\mathrm{i}\lambda 
_j}}{\cos(\frac{\lambda_j+\eta}{2}+\xi _+)},\\
& &\psi_2(\lambda _j,\xi_{-})=\frac{\cos(\frac{\lambda_j-\eta}{2}+\xi _-)e^{\mathrm{i}\lambda 
_j}}{\cos(\frac{\lambda_j+\eta}{2}-\xi _+)},
\end{eqnarray}
Case (I,II)
\begin{eqnarray}
& &\zeta_1(k_{i},\xi_{+})=\frac{\sin(\frac{k_i+\eta}{2}+\xi _+)}
{\sin(\frac{k_i-\eta}{2}-\xi _+)},\,\,
\zeta_2(k_i,\xi_{-})=-e^{\mathrm{i}k_i},\\
& &
\psi _1(\lambda _j,\xi_{+})=1, \,\,
\psi_2(\lambda _j,\xi_{-})=\frac{\cos(\frac{\lambda_j-\eta}{2}+\xi _-)e^{\mathrm{i}\lambda 
_j}}{\cos(\frac{\lambda_j+\eta}{2}-\xi _+)},
\end{eqnarray}
Case (II,I)
\begin{eqnarray}
& &\zeta_1(k_{i},\xi_{+})=-\frac{\cos(\frac{k_i}{2}-\xi _+)}
{\cos(\frac{k_i}{2}+\xi _+)},\,\,
\zeta_2(k_i,\xi_{-})=\frac{\sin(\frac{k_i+\eta}{2}+\xi _-)}
{\sin(\frac{k_i-\eta}{2}-\xi _-)},\\
& &
\psi _1(\lambda _j,\xi_{+})=\frac{\cos(\frac{\lambda_j-\eta}{2}-\xi _+)e^{-\mathrm{i}\lambda 
_j}}{\cos(\frac{\lambda_j+\eta}{2}+\xi _+)},\,\,
\psi_2(\lambda _j,\xi_{-})=1.
\end{eqnarray}
Above $M$ is the number of electrons with spin down and $n$ is the
total number of the electrons. Considering in addition the relations (\ref{H-DTM-1}) and
(\ref{H-DTM-2}) one can find the energy spectrum for all cases \\
Case (I,I)
\begin{equation}
E=-\frac{\cosh\beta}{c_1}-\frac{\cosh\beta}{c_2}+4\sum ^{n}_{i=1}\cos k_i,
\end{equation}
Case (II,II)
\begin{equation}
E=\exp\left(\beta \right)(c_1+c_2)+4\sum ^{n}_{i=1}\cos k_i,
\end{equation}
Case (I,II)
\begin{equation}
E=\exp\left(\beta \right)c_1-\frac{\cosh\beta}{c_2}+4\sum ^{n}_{i=1}\cos
k_i,
\end{equation}
Case (II,I)
\begin{equation}
E=-\frac{\cosh\beta}{c_1}+\exp\left(\beta \right)c_2+4\sum ^{n}_{i=1}\cos
k_i.
\end{equation}

\section{Conclusion and discussion}
\label{sec5}

In summary, we have analysed in detail the open
boundary conditions for the Bariev chain with special
boundary conditions. Two classes
of boundary $K_{+}$-matrices have been obtained
by solving the RE. It has been found that these two classes of
solutions of the RE lead to four kinds of integrable boundary fields
for the charge and spin degrees of freedom
separately. Through the nesting procedure, we have
diagonalized exactly the two level transfer matrices with four
kinds of possible boundary fields.  The eigenvalues of the transfer
matrices, the energy spectrum and Bethe ansatz equations have been
derived for all cases. The boundary fields indeed contribute
nontrivially 
to the ground-state properties as well as the low-lying spectrum. The
functions $\zeta_a(k_{i},\xi_{\pm })$ and $\psi _a(\lambda _j,\xi
_{\pm})$ will contribute nontrivial phase-shift factors to the density
of the charge rapidity and spin rapidity. We notice that the Bethe
equations (\ref{Bethe1}) and (\ref{Bethe2}) would be reduced to the
purely doubling of the ones for the 1D Bariev model with periodic BC
\cite{Mar}\cite{zhou2} if the boundary parameters $\xi _{\pm }\rightarrow
\infty$. 
These boundary parameters $\xi _{\pm }$, denoting the impurity strength too,
might change the band filling, the boundary surface energy, the
mesoscopic effects as well.  The results obtained provide us with a
basis to investigate the surface critical properties,
and correlation functions \cite{QISM2,FTP,corr}
for the resulting model.

\begin{ack}
X.W.G., A.F. and I.R. thank CNPq (Conselho Nacional de Desenvolvimento
Cient\'{\i}fico e Tecnol\'ogico) for financial support.
J.L. thanks the Australian Research Council for support. H.Q.Z. 
acknowledges the support from the NNSF of China.
\end{ack}
\appendix
\renewcommand{\thesection}{Appendix~\Alph{section}.}
\renewcommand{\theequation}{\Alph{section}.\arabic{equation}}
\section{The Boltzman weights and the boundary \protect\boldmath{$K_{\pm}$}-matrices}
Before practising our ansatz for fixing the boundary $K_{\pm }(u)$-matrix, we 
display the Boltzmann weights of the $R$-matrix (\ref{R}), which
appeared first in \cite{zhou}
\begin{eqnarray}
& &\rho _2=\frac{\sqrt{1+h^2u^2}\sqrt{1+h^2v^2}}{1+uvh^2}\rho _1,\,\,
\rho _3=\frac{(u-v)h}{1+uvh^2}\rho _1,\nonumber\\
& &\rho _4=\frac{\sqrt{1+h^2u^2}\sqrt{1+h^2v^2}\sqrt{1+u^2}\sqrt{1+v^2}}{(1+uv)(1+uvh^2)}\rho 
_1,\nonumber\\
& &\rho _5=\frac{h\sqrt{1+h^2v^2}\sqrt{1+u^2}(u-v)}{(1+uv)(1+uvh^2)}\rho _1,\nonumber\\
& &\rho _6=\frac{\sqrt{1+h^2v^2}\sqrt{1+u^2}(u-v)}{(1+uv)(1+uvh^2)}\rho _1,\,\,
\rho _7=\left[1+\frac{h^2(u-v)^2}{(1+uv)(1+uvh^2)}\right]\rho_1,\nonumber \\
& &\rho _8=\frac{\sqrt{1+u^2}\sqrt{1+v^2}}{1+uv}\rho _1,\,\, 
\rho _9=\frac{(u-v)(u-h^2v)}{(1+uv)(1+uvh^2)}\rho _1,\nonumber\\
& &\rho _{10}=\left[1+\frac{(u-v)^2}{(1+uv)(1+uvh^2)}\right]\rho_1,\,\,
\rho _{11}=\frac{(u-v)h}{1+uv}\rho _1,\nonumber\\
& &\rho _{12}=\frac{h\sqrt{1+h^2u^2}\sqrt{1+v^2}(u-v)}{(1+uv)(1+uvh^2)}\rho _1,\,\,\rho 
_{13}=\frac{\sqrt{1+h^2u^2}\sqrt{1+v^2}(u-v)}{(1+uv)(1+uvh^2)}\rho _1,
\nonumber\\
& &\rho _{14}=\frac{(u-v)(h^2u-v)}{(1+uv)(1+uvh^2)}\rho _1,\,\,
\rho _{15}=\frac{h(u-v)^2}{(1+uv)(1+uvh^2)}\rho _1,\nonumber
\end{eqnarray}
As mentioned in
section 2, due to the absence of the isomorphism between $K_{\pm
}$-matrices, we have to solve the RE (\ref{RE1}) (\ref{RE2})
separately to fix them.  Substituting $K_{-}(u)$ (\ref{BK}) into the
RE (\ref{RE1}), we find, at first glance, that the RE (\ref{RE1}) involve two
variables $u$ and $v$ which make the functional equations
involving $Ka_{-}(u),\, a=1,\ldots,4$ much more
complicated. However, if we observe the structures of the
$R$-matrix  and the RE (\ref{RE1}), we may pick up some
simpler functional equations of the RE which allow us to separate
$Ka_{-}(u),Ka_{-}(v) $ into the following forms
\begin{eqnarray}
\frac{K1_-(v)}{K2_-(v)}& = & 
\frac{\rho_2(u,v)\rho_3(v,-u)K1_-(u)+\rho_3(u,v)\rho_2(v,-u)K2_-(u)}{\rho_2(u,-v)\rho_3(-v,-u)K1
_-(u)+\rho_3(u,-v)\rho_2(-v,-u)K2_-(u)},\nonumber\\
\frac{K1_-(v)}{K3_-(v)}& = & 
\frac{\rho_2(u,v)\rho_3(v,-u)K1_-(u)+\rho_3(u,v)\rho_2(v,-u)K2_-(u)}{\rho_2(u,-v)\rho_3(-v,-u)K1
_-(u)+\rho_3(u,-v)\rho_2(-v,-u)K2_-(u)},\nonumber\\
\frac{K3_-(v)}{K4_-(v)}& = & 
\frac{\rho_{11}(u,-v)\rho_8(-v,-u)K3_-(u)+\rho_8(u,-v)\rho_{11}(-v,-u)K4_-(u)}{\rho_{11}(u,v)\rho_8(v,-u)K3_-(u)+\rho_8(u,v)\rho_2(v,-u)K4_-(u)},\nonumber
\end{eqnarray}
Substituting the Boltzmann weights of the $R$-matrix  into
above equations, and analying the consistant conditions of these
equations, we can easily conclude an ansatz
\begin{eqnarray}
K1_-(u)&=& (c_1+c_2u)(c_3+c_4u)(c_5+c_6u),\label{stru1}\\ 
K2_-(u)&=& (c_1-c_2u)(c_3+c_4u)(c_5+c_6u),\label{stru2}\\
K3_-(u)&=& (c_1+c_2u)(c_3-c_4u)(c_5+c_6u),\label{stru3}\\
K4_-(u)&=& (c_1+c_2u)(c_3-c_4u)(c_5-c_6u),\label{stru4}
\end{eqnarray}
with minimal coefficients $c_i$ to be determined.
Running the RE again with this ansatz
(\ref{stru1})--(\ref{stru4}), it is easily found that only one
coefficient is free. Thus the two classes of boundary $K_-$-matrices
can be  immediately chosen as the forms (\ref{Km1}) and (\ref{Km2}).

Simlarly, substituting $K_+$-matrix (\ref{BK}) into the RE (\ref{RE2}) we have
\begin{eqnarray}
\frac{K1_+(v)}{K2_+(v)}& = & 
\frac{\rho_2(v,u)\bar{\rho}_{16}(v,-u)K1_+(u)+\rho_3(v,u)\bar{\rho}_3(v,-u)K2_+(u)}{\rho_3(-u,-v
)\bar{\rho}_2(u,-v)K1_+(u)+\rho_2(-u,-v)\bar{\rho}_{16}(u,-v)K2_+(u)},\nonumber\\
\frac{K1_+(v)}{K3_+(v)}& = & 
\frac{\rho_2(v,u)\bar{\rho}_{16}(v,-u)K1_+(u)+\rho_3(v,u)\bar{\rho}_2(v,-u)K3_+(u)}{\rho_3(-u,-v
)\bar{\rho}_3(u,-v)K1_+(u)+\rho_2(-u,-v)\bar{\rho}_{16}(u,-v)K3_+(u)},\nonumber\\
\frac{K3_+(v)}{K4_+(v)}& = & 
\frac{\rho_8(-u,-v)\bar{\rho}_{15}(u,-v)K3_+(u)+\rho_{11}(-u,-v)\bar{\rho}_7(u,-v)K4_+(u)}{\rho_
{11}(v,u)\bar{\rho}_{9}(v,-u)K3_+(u)+\rho_{8}(v,u)\bar{\rho}_{15}(v,-u)K4_+(u)},\nonumber
\end{eqnarray}
here we first  have to determine $\bar{\rho}$'s from the relation (\ref{tildeR1}) and 
(\ref{tildeR2}), i.e.
\begin{eqnarray}
\bar{\rho }_2(x,y) & = 
&\frac{(1+h^2xy)^2(1+xy)\sqrt{1+h^2y^2}\sqrt{1+h^2x^2}}{(x-y)^2(h^2x-y)(h^2y-x)},\nonumber\\
\bar{\rho }_3(x,y) & = 
&\frac{(1+h^2xy)^2(1+xy)\sqrt{1+h^2y^2}\sqrt{1+h^2x^2}}{(x-y)^2(h^2x-y)(h^2y-x)h^2},\nonumber\\
\bar{\rho }_7(x,y) & = 
&\frac{(1+h^2xy)^2(1+xy)^2\sqrt{1+y^2}\sqrt{1+x^2}}{(x-y)^2(h^2x-y)(h^2y-x)},\nonumber\\
\bar{\rho }_9(x,y) & = 
&\frac{(1+h^2xy)^2(1+xy)^2h^2\sqrt{1+y^2}\sqrt{1+x^2}}{(x-y)^2(h^2x-y)(h^2y-x)},\nonumber\\
\bar{\rho }_{15}(x,y) & = &\frac{(1+h^2xy)(1+xy)^2}{(x-y)^2(xh^2-y)},\nonumber\\
\bar{\rho }_{16}(x,y) & = &\frac{(1+h^2xy)^2(1+xy)}{(x-y)^2(yh^2-x)h}.\nonumber
\end{eqnarray}
With the help of the new Boltzmann weights, the consistency of above 
equations indeed  can be able to lead to an ansatz
\begin{eqnarray}
K1_+(u)&=& (c_1+c_2u)(c_3+c_4uh^2)(c_5+c_6u),\label{stru12}\\ 
K2_+(u)&=& (c_2u-h^2c_1)(c_4uh^2+c_3)(c_5+c_6u),\label{stru22}\\
K3_+(u)&=& (c_1+c_2u)(c_4u-c_3)(c_5+c_6u),\label{stru32}\\
K4_+(u)&=& (c_1+c_2u)(c_4u-c_3)(c_6u-c_5h^2). \label{stru42}
\end{eqnarray}
Running the second RE (\ref{RE2}) again, the solutions (\ref{Kp1}) and 
(\ref{Kp2}) would be fixed.
\section{Useful commutation relations}

For factorizing the transfer matrix (\ref{DTM}) acting on the
pseudo-vacuum state, we need the following commutation relations
\begin{eqnarray}
C_1(u)\bar{B}_1(u)& = &
 \frac{\rho_2(u,-u)}{\rho_1(u,-u)}\left[\bar{B}(u)B(u)
 -A_{11}(u)\bar{A}_{11}(u)\right],\\
\bar{C}_1(u){B}_1(u)& = &
 \frac{\rho_2(u,-u)}{\rho_1(u,-u)}\left[{B}(u)\bar{B}(u)
 -\bar{A}_{11}(u){A}_{11}(u)\right],\\
C_2(u)\bar{B}_2(u)& = &
 \frac{\rho_2(u,-u)}{\rho_1(u,-u)}\left[\bar{B}(u)B(u)
 -A_{22}(u)\bar{A}_{22}(u)\right],\\
\bar{C}_2(u){B}_2(u)& = &
 \frac{\rho_2(u,-u)}{\rho_1(u,-u)}\left[{B}(u)\bar{B}(u)
 -\bar{A}_{22}(u){A}_{22}(u)\right],\\
C_3(u)\bar{F}(\lambda)& = &
\frac{\rho_4(u,-u)}{\rho_1(u,-u)}\left[\bar{B}(u)B(u)-
 D(u)\bar{D}(u)\right]\nonumber\\
& &
 -\frac{\rho_2(u,-u)}{\rho_1(u,-u)}\left[{C}_4(u)\bar{E}_1(u)+
                   {C}_5(u)\bar{E}_2(u)\right],\\
C_3(u)\bar{F}(u)& = &
\frac{\rho_4(u,-u)}{\rho_2(u,-u)}\bar{C_1}(u)B_1(u)+
\frac{\rho_8(u,-u)}{\rho_2(u,-u)}\left[\bar{A}_{11}(u)A_{11}(u)-D(u)\bar{D}(u)\right]\nonumber\\
& &
 -\frac{\rho_1(u,-u)}{\rho_2(u,-u)}{C}_4(u)\bar{E}_1(u)-
\frac{\rho_7(u,-u)}{\rho_2(u,-u)}{C}_5(u)\bar{E}_2(u),\\
C_3(u)\bar{F}(u)& = &
\frac{\rho_4(u,-u)}{\rho_2(u,-u)}\bar{C_2}(u)B_2(u)+
\frac{\rho_8(u,-u)}{\rho_2(u,-u)}\left[\bar{A}_{22}(u)A_{22}(u)-D(u)\bar{D}(u)\right]\nonumber\\
& &
 -\frac{\rho_{10}(u,-u)}{\rho_{2}(u,-u)}{C}_4(u)\bar{E}_1(u)-
\frac{\rho_1(u,-u)}{\rho_2(u,-u)}{C}_5(u)\bar{E}_2(u),
\end{eqnarray}
which can be derived directly from (\ref{YBA}). 
In addition, one also can show the relations
$C_i(\lambda)\bar{B}_a(\lambda)=0$ for $i\neq j$, $i=1,2,3$, and
$a=1,2$. Using these commutation relations and after lengthy algebra, 
one may manage to present the factorized 
forms of (\ref{fact-m}) and (\ref{DTM-E}).

In order to check the eigenvalue problem of two-particle excitation, the
following commutation equations are necessary
\begin{eqnarray}
& &\rho _1(u,v)\rho _{14}(v,-u)\tilde{F}(u)\tilde{B}(v)=
\rho _{14}(u,-v)\rho _{4}(-v,-u)\tilde{F}(v)\tilde{B}(u)\nonumber\\
& &
+\rho _{12}(-v,-u)\left[\rho_3(u,-v)\tilde{B}_2(v)\tilde{B}_1(u)
+\rho_{12}(u,-v)\tilde{F}_2(v)\tilde{A}_{11}(u)\right]\nonumber\\
& &
+\rho _{13}(-v,-u)\left[\rho_3(u,-v)\tilde{B}_1(v)\tilde{B}_2(u)
+\rho_{13}(u,-v)\tilde{F}_2(v)\tilde{A}_{22}(u)\right]\nonumber\\
& &
+\rho _{14}(-v,-u)\left[\rho_1(u,-v)\tilde{B}(v)\tilde{F}(u)
+\rho_{2}(u,-v)\tilde{B}_1(v)\tilde{E}_{1}(u)\right.\nonumber\\
& &
\left.+\rho_2(u,-v)\tilde{B}_2(v)\tilde{E}_2(u)
+\rho_{4}(u,-v)\tilde{F}(v)\tilde{D}(u)\right],\\
& &
\rho _{9}(u,v)\rho _{14}(v,-u)\tilde{D}(u)\tilde{F}(v)+\nonumber\\
& &
\rho _{5}(u,v)\left[\rho_{12}(v,-u)\tilde{A}_{11}(u)\tilde{F}(v)
+\rho_{11}(v,-u)\tilde{E}_1(u)\tilde{E}_{2}(v)\right]\nonumber\\
& &
+\rho _{6}(u,v)\left[\rho_{13}(v,-u)\tilde{A}_{22}(u)\tilde{F}(v)
+\rho_{11}(v,-u)\tilde{E}_2(u)\tilde{E}_{1}(v)\right]\nonumber\\
& &
+\rho _{4}(u,v)\left[\rho_{4}(v,-u)\tilde{B}(u)\tilde{F}(v)
+\rho_{8}(v,-u)\tilde{B}_1(u)\tilde{E}_{1}(v)\right.\nonumber\\
& &
\left.+\rho_{8}(v,-u)\tilde{B}_2(u)\tilde{E}_{2}(v)
+\rho_{1}(v,-u)\tilde{F}(u)\tilde{D}(v)\right]\nonumber\\
& &=\rho_4(u,-v)\rho_1(-v,-u)\tilde{B}(v)\tilde{F}(u)+
\rho_1(u,-v)\rho_1(-v,-u)\tilde{F}(v)\tilde{D}(u)\nonumber\\
& &
+\rho _{8}(u,-v)\left[\rho_1(-v,-u)\tilde{B}_1(v)\tilde{E}_1(u)
+\rho_{1}(-v,-u)\tilde{B}_2(v)\tilde{E}_{2}(u)\right],\\
& &
\rho _{3}(u,v)\left[\rho _{3}(v,-u)\tilde{A}_{11}(u)\tilde{F}(v)+
\rho_{13}(v,-u)\tilde{E}_{1}(u)\tilde{E}_2(v)\right]\nonumber\\
& &
+\rho _{2}(u,v)\left[\rho_{2}(v,-u)\tilde{B}(u)\tilde{F}(v)
+\rho_{1}(v,-u)\tilde{B}_1(u)\tilde{E}_{1}(v)\right.\nonumber\\
& &
\left.+\rho_{10}(v,-u)\tilde{B}_2(u)\tilde{E}_{2}(v)
+\rho_{8}(v,-u)\tilde{F}(u)\tilde{D}(v)\right]\nonumber\\
& &=\rho_{11}(-v,-u)\left[\rho_{13}(u,-v)\tilde{B}_2(v)\tilde{B}_1(u)+
\rho_{11}(u,-v)\tilde{F}(v)\tilde{A}_{11}(u)\right]\nonumber\\
& &
+\rho _{8}(-v,-u)\left[\rho_2(u,-v)\tilde{B}(v)\tilde{F}(u)
+\rho_{1}(u,-v)\tilde{B}_1(v)\tilde{E}_{1}(u)\right.\nonumber\\
& &
\left.+\rho_{10}(u,-v)\tilde{B}_2(v)\tilde{E}_2(u)
+\rho_{8}(u,-v)\tilde{F}(v)\tilde{D}(u)\right],\\
& &
\rho _{3}(u,v)\left[\rho _{3}(v,-u)\tilde{A}_{22}(u)\tilde{F}(v)+
\rho_{12}(v,-u)\tilde{E}_{2}(u)\tilde{E}_1(v)\right]\nonumber\\
& &
+\rho _{2}(u,v)\left[\rho_{2}(v,-u)\tilde{B}(u)\tilde{F}(v)
+\rho_{7}(v,-u)\tilde{B}_1(u)\tilde{E}_{1}(v)\right.\nonumber\\
& &
\left.+\rho_{1}(v,-u)\tilde{B}_2(u)\tilde{E}_{2}(v)
+\rho_{8}(v,-u)\tilde{F}(u)\tilde{D}(v)\right]\nonumber\\
& &=\rho_{11}(-v,-u)\left[\rho_{13}(u,-v)\tilde{B}_2(v)\tilde{B}_1(u)+
\rho_{11}(u,-v)\tilde{F}(v)\tilde{A}_{22}(u)\right]\nonumber\\
& &
+\rho _{8}(-v,-u)\left[\rho_2(u,-v)\tilde{B}(v)\tilde{F}(u)
+\rho_{7}(u,-v)\tilde{B}_1(v)\tilde{E}_{1}(u)\right.\nonumber\\
& &
\left.+\rho_{1}(u,-v)\tilde{B}_2(v)\tilde{E}_2(u)
+\rho_{8}(u,-v)\tilde{F}(v)\tilde{D}(u)\right],\\
& &
\rho _{3}(u,v)\left[\rho _{3}(v,-u)\tilde{A}_{12}(u)\tilde{F}(v)+
\rho_{12}(v,-u)\tilde{E}_{1}(u)\tilde{E}_1(v)\right]+\rho_2(u,v)\rho 
_{15}(v,-u)\tilde{B}_2(u)\tilde{E}_1(v)\nonumber\\
& &
=\rho_{11}(-v,-u)\left[\rho_{12}(u,-v)\tilde{B}_1(v)\tilde{B}_2(u)+
\rho_{11}(u,-v)\tilde{F}(v)\tilde{A}_{12}(u)\right]\nonumber\\
& &
+\rho_{15}(u,-v)\rho_8(-v,-u)\tilde{B}_2(v)\tilde{E}_1(u),\\
& &
\rho _{3}(u,v)\left[\rho _{3}(v,-u)\tilde{A}_{21}(u)\tilde{F}(v)+
\rho_{13}(v,-u)\tilde{E}_{2}(u)\tilde{E}_2(v)\right]+\rho_2(u,v)\rho 
_{15}(v,-u)\tilde{B}_1(u)\tilde{E}_2(v)\nonumber\\
& &
=\rho_{11}(-v,-u)\left[\rho_{12}(u,-v)\tilde{B}_1(v)\tilde{B}_1(u)+
\rho_{11}(u,-v)\tilde{F}(v)\tilde{A}_{21}(u)\right]\nonumber\\
& &
+\rho_{15}(u,-v)\rho_8(-v,-u)\tilde{B}_1(v)\tilde{E}_2(u).
\end{eqnarray}


\end{document}